\documentclass{SciPost_UpperCaseTitles}

\usepackage[bitstream-charter]{mathdesign}
\DeclareSymbolFont{usualmathcal}{OMS}{cmsy}{m}{n}
\DeclareSymbolFontAlphabet{\mathcal}{usualmathcal}

\fancypagestyle{SPstyle}{
\fancyhf{}
\lhead{\colorbox{scipostblue}{\bf \color{white} ~SciPost Physics }}
\rhead{{\bf \color{scipostdeepblue} ~Submission }}

\fancyfoot[C]{\textbf{\thepage}}
}

\usepackage{graphicx}
\usepackage{color}
\usepackage{verbatim}
\usepackage{mathtools}

\usepackage[normalem]{ulem}  

\usepackage{xcolor}
\hypersetup{
    colorlinks,
    linkcolor={red!50!black},
    citecolor={blue!50!black},
    urlcolor={blue!80!black}
}

\newcommand{\overbar}[1]{\mkern 3mu\overline{\mkern-3mu#1\mkern-3mu}\mkern 3mu}

\newcommand{\bl}{} 
\definecolor{darkgreen}{rgb}{0.09, 0.55, 0.3}

\definecolor{darkred}{rgb}{0.8, 0.10, 0.1}

\newcommand{\non}{\nonumber}
\newcommand{\bea}{\begin{eqnarray}}
\newcommand{\eea}{\end{eqnarray}}
\newcommand{\be}{\begin{equation}}
\newcommand{\ee}{\end{equation}}
\newcommand{\bes}{\begin{equation*}}
\newcommand{\ees}{\end{equation*}}
\newcommand{\bi}{\begin{itemize}}
\newcommand{\ei}{\end{itemize}}

\renewcommand{\vec}{\mathbf}

\newcommand{\rr}{\vec{r}}

\newcommand{\vn}{\vec{0}}
\newcommand{\ra}{\rangle}
\newcommand{\la}{\langle}

\newcommand{\up}{\uparrow}
\newcommand{\down}{\downarrow}

\newcommand{\Dr}{\mathcal{D}}

\newcommand{\Nr}{N} 

\newcommand{\Gex}{G_{\rm exact}}
\newcommand{\GtN}{G_\Nr} 
\newcommand{\Gtinf}{G_\infty} 
\newcommand{\tSig}{\Sigma} 
\newcommand{\tD}{D} 
\newcommand{\spin}{{ s}}

\renewcommand{\sout}[1]{\unskip}

\begin{document}

\pagestyle{SPstyle}

\begin{center}{\Large \textbf{\color{scipostdeepblue}{
        Physical and unphysical regimes
\vskip 0.1cm
        of self-consistent many-body perturbation theory
}}}\end{center}

\begin{center}
K. Van Houcke\textsuperscript{1},
E. Kozik\textsuperscript{2},
R. Rossi\textsuperscript{1,3,4,5},
Y. Deng\textsuperscript{6,7}, and
F. Werner\textsuperscript{8*}
\end{center}

\begin{center}
  {\bf 1} {\small Laboratoire de Physique de l'\'Ecole normale sup\'erieure, ENS - Universit\'e PSL, CNRS,
    \\Sorbonne Universit\'e, Universit\'e Paris Cité, 75005 Paris, France}
\\
{\bf 2} {\small Physics Department, King's College, London WC2R 2LS, United Kingdom}
\\
  {\bf 3}  {\small Center for Computational Quantum Physics, Flatiron Institute, 
    New York, NY 10010, USA}
  \\
    {\bf 4}  {\small CNRS, LPTMC, Sorbonne Université, 75005 Paris, France
    }  \\
  {\bf 5}  {\small Institute  of  Physics,    EPFL,    1015  Lausanne,    Switzerland} \\
{\bf 6}  {\small National Laboratory for Physical Sciences at Microscale and Department of Modern Physics, University of Science and Technology of China, Hefei, Anhui 230026, China}
\\
  {\bf 7}    {\small Shanghai Research Center for Quantum Science, Shanghai 201315, China}
  \\
    {\bf 8} {\small Laboratoire Kastler Brossel,
      ENS - Universit\'e PSL, CNRS,
      Coll\`ege de France,  Sorbonne Universit\'e, 75005 Paris, France}


    
* {\small werner@lkb.ens.fr}

\end{center}

\begin{center}
\today
\end{center}

\section*{\color{scipostdeepblue}{Abstract}}
{\boldmath
  \textbf{In the standard framework of self-consistent {many-body} perturbation theory,
  the skeleton series for the self-energy is truncated at a finite order $\Nr$ and plugged into the Dyson equation,
  which is then solved for
  the propagator $\GtN$. We consider two
examples of fermionic models,
the Hubbard atom at half filling and
  its zero space-time dimensional simplified version.
  First, we show that $\GtN$ converges when $\Nr\to\infty$ to a limit $G_\infty\,$, which coincides
with the exact physical propagator $\Gex$ 
at small enough coupling,
while 
$G_\infty \neq \Gex$
at strong coupling.
This follows from the findings of~\cite{KFG} and an additional subtle mathematical mechanism elucidated here.
Second, we demonstrate that it is possible to discriminate 
between the $\Gtinf=\Gex$ and $\Gtinf\neq\Gex$ regimes
thanks to a criterion
which does not require the knowledge of $\Gex\,$,
as proposed in \cite{ShiftedAction}.
}}

\vspace{10pt}
\noindent\rule{\textwidth}{1pt}
\tableofcontents
\noindent\rule{\textwidth}{1pt}

\section{Introduction}

Self-consistent 
perturbation theory is a particularly
elegant and powerful approach in quantum many-body physics \cite{MartinReiningCeperley,DupuisVol1,StefanucciVanLeeuwen}.
The single-particle propagator $G$ 
is expressed through the Dyson equation
\be
G^{-1} \ = \ G_0^{-1} \ -\ \Sigma
\label{eq:dyson}
\ee
in terms of the non-interacting propagator $G_0$ and the self-energy $\Sigma$,
which itself is formally expressed in terms of $G$ through
the skeleton series,
\be
\Sigma \ =\ \Sigma_{\rm bold}[G] \ \equiv \ \sum_{n=1}^\infty \  \Sigma_{\rm bold}^{(n)}[G]
\label{eq:sig_bold}
\ee
where $\Sigma_{\rm bold}^{(n)}[G]$ is the sum of all
skeleton self-energy
Feynman diagrams of order $n$ (these diagrams are built with bold propagator lines representing $G$, and remain connected if one cuts one or two $G$-lines).

The standard procedure for solving Eqs.~(\ref{eq:dyson},\ref{eq:sig_bold}) 
is to truncate the skeleton series at a finite order $\Nr$,
and to look for the solution $\GtN$ of the self-consistency equation\footnote{We assume that the solution $\GtN$ of (\ref{eq:bold_N}) is unique, or at least that there is no difficulty in identifying a unique 
potentially physical solution ({\it e.g.}, by
starting from the weakly interacting limit
where $\GtN \to G_0$,
and following the solution as a function of interaction strength).}
\be
\GtN^{-1} \ = \ G_0^{-1} \ - \ \Sigma_{\rm bold}^{(\leq \Nr)}[\GtN]
\label{eq:bold_N}
\ee
with
\be
\Sigma_{\rm bold}^{(\leq \Nr)} \ \, := \ \, \sum_{n=1}^\Nr \ \Sigma_{\rm bold}^{(n)}\,.
\non 
\ee
The natural expectation is that
one obtains the exact propagator by sending the truncation order to infinity:
$\GtN \to \Gex$ for $\Nr\to\infty$.

However,
as was discovered in~\cite{KFG},
the series $\Sigma_{\rm bold}^{(\leq \Nr)}[\Gex]$
can converge when $\Nr\to\infty$ to a result which
differs from the exact physical self-energy $\Sigma_{\rm exact} = G_0^{-1} - \Gex^{-1}$\,.
This
misleading convergence 
phenomenon
was observed 
for three fermionic textbook models ---Hubbard atom, Anderson impurity model, and half-filled 2D Hubbard model---
in a region of the parameter space
(at and around half filling, at strong interaction and low temperature).
$\Gex$ was computed with a numerically exact
quantum Monte Carlo method,
and the skeleton series
was evaluated
up to $\Nr = 6$ or 8 by diagrammatic Monte Carlo~\cite{VanHoucke1}.
Numerous works~\cite{reining_2solutions,Rossi_Werner_0+0dim,ShiftedAction,
SchaeferDivergences2016,ReiningHubbat,GunnarsonBreakdownPRL,ParcolletMultival,LinLin,ToschiDivergences2018,KimMultival}
have studied
various aspects of the problem found in~\cite{KFG},
as well as the related divergences of irreducible vertices (\cite{SchaeferDivergences2016,GunnarsonBreakdownPRL,ParcolletMultival,ToschiDivergences2018,SchaeferDivergentPrecursors,GunnarsonParquetDCA2016,RohringerRevue,ToschiDvg2020,AdlerNonperturbative} and Refs. therein).
In particular, Ref.~\cite{Rossi_Werner_0+0dim} introduced an exactly solvable toy model, which has the structure of a fermionic model in zero space-time dimensions,
and features the misleading convergence problem of~\cite{KFG}\,,
as well as the related multivaluedness of the Luttinger-Ward functional also discovered in~\cite{KFG}.

In this article, we study the consequences of this problem
for the 
sequence $\GtN$,
which is the crucial question in the most relevant cases where $\Gex$ is unknown.
For the toy model of~\cite{Rossi_Werner_0+0dim}, we find that 
$\GtN$
converges when $\Nr\to\infty$ to
a limit $\Gtinf$ which differs from $\Gex$ at strong coupling;
for the Hubbard atom, our numerical data strongly indicate that
such misleading convergence of the sequence $\GtN$ also occurs at large coupling and half filling.
This misleading convergence of $\GtN$ is the first result reported in this article.
Secondly, we present data, again for the toy model of~\cite{Rossi_Werner_0+0dim} and for the Hubbard atom, demonstrating that a criterion proposed in~\cite{ShiftedAction} enables one to discriminate between the $\Gtinf\neq\Gex$ and $\Gtinf=\Gex$ regimes without using the knowledge of $\Gex$.

The misleading convergence of $G_N$ reported here is a 
non-trivial fact.
 It comes from a subtle mathematical mechanism (as we will see), and
 does not merely follow from the misleading convergence of 
  \,$\Sigma_{\rm bold}^{(\leq \Nr)}[\Gex]$
  discovered in~\cite{KFG}.
  Indeed, a naive reasoning would suggest that if the misleading convergence of 
  \,$\Sigma_{\rm bold}^{(\leq \Nr)}[\Gex]$ takes place,
then  $G_N$ should not converge at all.\footnote{The naive reasoning goes as follows:
If $G_N$ would converge to some $G_\infty$ for $N\to\infty$,
then, from~(\ref{eq:bold_N}), one can expect $G_\infty^{-1} = G_0^{-1}-\Sigma_{\rm bold}[G_\infty]$, and hence, assuming unicity of the solution of the Dyson equation, $G_\infty=\Gex$.
Thus $\Sigma_{\rm bold}[\Gex] = \Sigma_{\rm exact}$, in contradiction with the misleading convergence of $\Sigma_{\rm bold}^{(\leq \Nr)}[\Gex]$.}

We restrict here to the scheme~(\ref{eq:dyson},\ref{eq:sig_bold})
where $G$ is the only bold element
(as in, {\it e.g.}, Ref.~\cite{CarlstromLineNodeSemimetal}). Nevertheless,
our findings may also be relevant to
other schemes containing additional bold elements,
such as
a bold interaction line $W$,
or a bold pair propagator line $\Gamma$.
The scheme built with $G$ and $W$
is natural for Coulomb interactions,
and is widely used for solids and molecules with a truncation order $\Nr=1$
(the $GW$ approximation) 
and sometimes with $\Nr=2$ (see, {\it e.g.}, Refs.~\cite{Gunnarson_GW_Revue,SimonsMolecules,Leeuwen_GW2_mol,Zgid_GW2}),
while for several paradigmatic lattice models,
bold diagrammatic Monte Carlo (BDMC) made it possible to reach larger $\Nr$ and claim a small residual truncation error~\cite{IgorCoulombPhonon,IgorDirac,IgorHaldane,SimonsHydrogenChain}.
The scheme built with $G$ and $\Gamma$
is natural
for contact interactions;
 truncation at order $\Nr=1$
then corresponds to the self-consistent T-matrix approximation~\cite{Haussmann_Z_Phys,Haussmann_PRB,HaussmannZwergerThermo},
and precise large-$\Nr$  results were obtained by BDMC in the normal phase of the Hubbard model \cite{DengEmergentBCS,SimkovicEmergent} and of
the unitary Fermi gas \cite{VanHouckeEOS,RossiEOS,RossiContact}.
Other BDMC results were obtained
for models of coupled electrons and phonons, where it is natural to introduce a bold phonon propagator 
\cite{MishchenkoProkofevPRL2014,IgorCoulombPhonon},
and for frustrated spins~\cite{KulaginPRL,HuangPyro,WangCaiSpins}.
Schemes containing three- or four-point bold vertices were also employed, to construct extensions of dynamical mean-field theory~\cite{AyralFunctional,RohringerRevue}.

\section{Zero space-time dimensional toy-model}


\subsection{Definitions and reminders}

We begin with some reminders from~\cite{Rossi_Werner_0+0dim} (see \cite{Rossi_Werner_0+0dim} for the derivations).
While fermionic many-body problems can be represented by a functional integral over Grassmann {\it fields}, which depend on $d$ space coordinates and one imaginary time coordinate~\cite{NegeleOrland,DupuisVol1},
in
this simplified toy model
the Grassmann fields
are replaced with Grassmann {\it numbers}
$\varphi_\spin$ and $\overbar{\varphi}_\spin$
that do not depend on anything,
apart from a spin index
 $\spin\in\{\uparrow,\downarrow\}$.
The partition function, the action and the propagator are then defined by
\begin{equation}
Z= \int \left(\prod_{\spin} d\mkern-2.5mu\varphi_\spin  \  d\mkern-2mu\overbar{\varphi}_\spin\right)\; e^{-S} 
\nonumber
\end{equation}
\be
S 
\ = \ -\mu \, \sum_\spin  \overbar{\varphi}_\spin  \varphi_\spin
\ + \ U \ \, \overbar{\varphi}_\uparrow \varphi_\uparrow \overbar{\varphi}_\downarrow \varphi_\downarrow
\nonumber
\ee
\be
G
\ =
\ -\,\frac{1}{Z}\ \int \left(\prod_\spin d\mkern-2.5mu\varphi_\spin \  d\mkern-2mu\overbar{\varphi}_\spin\right)\  \, \varphi_{\spin'}\, \overbar{\varphi}_{\spin'}\ \, e^{-S} \,, 
\nonumber
\ee
the dimensionless parameters
$\mu$ and $U$
being the analogs of
chemical potential and interaction strength.
Since $G$ is spin-independent, we omit its spin index.
We restrict for convenience to $\mu>0$
  (changing the sign of $\mu$ essentially amounts to the change of variables $\varphi \leftrightarrow \overbar{\varphi}$\,)
and to $U<0$ (as in \cite{Rossi_Werner_0+0dim}\,).

The coefficients of the skeleton series have the analytical expression
\be
\Sigma_{\rm bold}[G] \ = \ \sum_{n=1}^\infty\,
{a_n}\,G^{2 n - 1} U^n
\ \ \ \ \ \  {\rm with}\ \ \ \ \ \ {a_n} = \frac{(-1)^{n-1} (2n-2)!}{n! (n-1)!}.
\nonumber
\ee
It is convenient to work with {rescaled variables,
 multiplying propagators with $\sqrt{|U|}$
  and dividing self-energies with the same factor,} 
\be
g \coloneqq G\sqrt{|U|}\ ,
\ \ \ \ \ \ 
\sigma \coloneqq \Sigma / \sqrt{|U|}\ .
\label{eq:rescaled}
\ee
The rescaled skeleton series is then given by
\be
\sigma_{\rm bold}(g) \ = \ \sum_{n=1}^\infty \ \sigma_{\rm bold}^{(n)}(g)
\ \ \ \ \ \ 
{\rm with}
\ \ \ \ \ \ 
\sigma_{\rm bold}^{(n)}(g) = {a_n} (-1)^n g^{2n-1}
\nonumber
\ee
and accordingly
$\sigma_{\rm bold}^{(\leq\Nr)}(g) \, \equiv \ \sum_{n=1}^\Nr \sigma_{\rm bold}^{(n)}(g)$\,.

The exact self-energy and 
propagator are given by
    \bea
    \sigma_{\rm  exact}(g_0) &=& -g_0 \nonumber
    \\
    g_{\rm exact}(g_0) &=& \frac{g_0}{1+g_0^2}  \nonumber    
    \eea
in terms of the rescaled free propagator $g_0 := \sqrt{|U|} \, G_0  = \sqrt{|U|}/\mu$\,.

\begin{figure}[h!]
  \begin{center}
    \includegraphics[width=0.58\columnwidth]{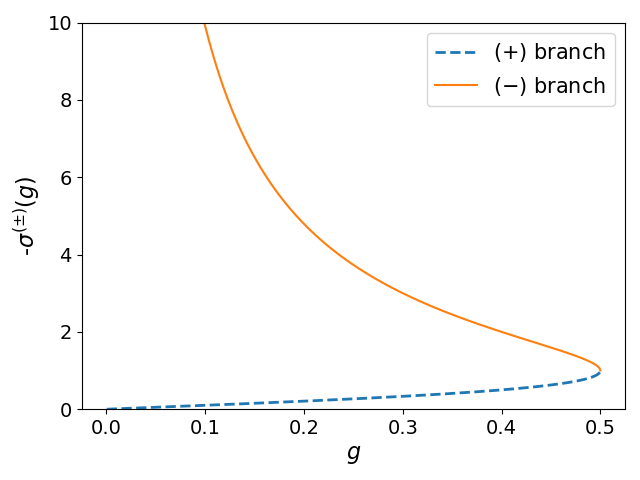}
    \caption{The two branches of the self-energy as a function of the full propagator, for the toy model in zero space-time dimensions.
      The skeleton series converges up to $g=1/2$ and coincides with the $(+)$ branch:
       $\sigma_{\rm bold}(g) = \sigma^{(+)}(g)$ for $g \leq 1/2$.
      \label{fig:sigma_pm}}
\end{center}
  \end{figure}

    If one evaluates the skeleton series at the exact $G$, one obtains the correct physical self-energy for $|U|<\mu^2$ and an incorrect result for $|U|>\mu^2$.
This is directly related to the fact that
    the self-energy functional (which reduces to a function in this toy model)
    has two branches,
    \be
    \sigma^{(\pm)}(g) = \frac{-1\pm\sqrt{1-4 g^2}}{2g}
    \label{eq:sig_pm}
\ee
as represented in Fig.~\ref{fig:sigma_pm}
(this corresponds to the derivative of the two branches of the Luttinger-Ward functional, restricting to spin-independent $G$ for simplicity).
The physical branch is the $(+)$ branch for $g_0<1$, and the $(-)$ branch for $g_0>1$; {\it i.e.},
  $\sigma_{\rm  exact}(g_0) \, = \,
    \sigma^{({\rm sign}(1-g_0))}(g_{\rm exact}(g_0))$\,.
On the other hand, the skeleton series, evaluated at the exact physical propagator, always converges to the $(+)$ branch;
{\it i.e.},  $\sigma_{\rm bold}(g_{ \rm exact}(g_0)) \, = \, \sigma^{(+)}(g_{\rm exact}(g_0))$ for all $g_0>0$.

Note that $\sigma_{\rm bold}(g)$ is the expansion of $\sigma^{(+)}(g)$ in powers of $g$, and
thus from (\ref{eq:sig_pm}) the convergence radius of the series $\sigma_{\rm bold}(g)$ is 1/2.


\vskip -4cm

\subsection{Limit of the skeleton sequence}

We now go beyond Ref.~\cite{Rossi_Werner_0+0dim} and study the ``skeleton sequence'' $\GtN$ defined by Eq.~(\ref{eq:bold_N}).
Rescaling variables as in (\ref{eq:rescaled}),
  in particular setting $g_\Nr\coloneqq \GtN\sqrt{|U|}$\,,
 Equation~(\ref{eq:bold_N}) becomes
\be
\frac{1}{g_\Nr} \ =\ \frac{1}{g_0} \ - \ \sigma_{\rm bold}^{(\leq\Nr)}(g_\Nr)\,.
\label{eq:gN}
\ee
This equation is readily solved for $g_\Nr$ numerically:
The solutions are roots of a polynomial of order $2\Nr$,
and we observe that there is a unique real positive root,
which we take to be $g_\Nr$
(recall that the exact $g$ is always real and positive);
alternatively,
we solved Eq.~(\ref{eq:gN}) by iterations
(with a damping procedure described in the next section), 
and we found convergence to this same $g_\Nr$.
We find that
\bi
\item
  for $g_0<1$, \ \ $g_\Nr\underset{\Nr\to\infty}{\longrightarrow}\ g_{\rm  exact}(g_0)$
\item
  for $g_0>1$, \ \ $g_\Nr\underset{\Nr\to\infty}{\longrightarrow} \ g_\infty \ \neq \ g_{\rm  exact}(g_0)$
  \ei {\it i.e.}, the skeleton sequence converges to the correct physical result below a critical coupling strength, and to an unphysical result above it. 

\begin{figure}[h!]
    \begin{center}
       \includegraphics[width=0.75\columnwidth]{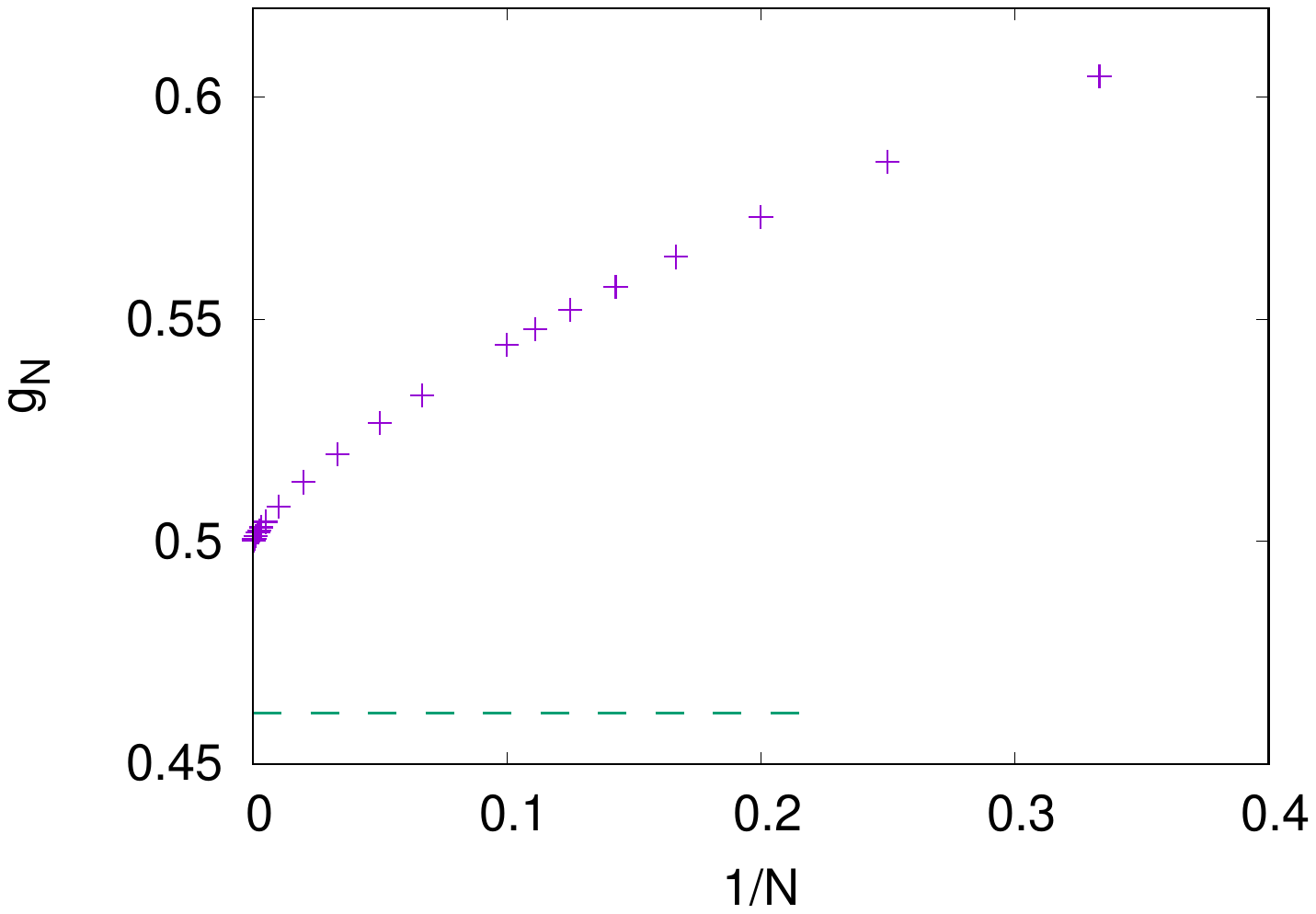}
\vskip -0.7cm
       \caption{
         {\it Illustrative example of
misleading convergence
of the skeleton sequence for the toy model.}
The rescaled propagator $g_N$, obtained from the self-consistency equation with the skeleton series truncated at order $N$, converges
for $N\to\infty$ to the limit 0.5, which differs from the exact
result (dashed line).
This happens when the rescaled free propagator $g_0>1$ (here, $g_0=1.5$).       \label{fig:cvg_wrong}}
         \end{center}
       \end{figure}

\begin{figure}[h!]
    \begin{center}
       \includegraphics[width=0.9\columnwidth]{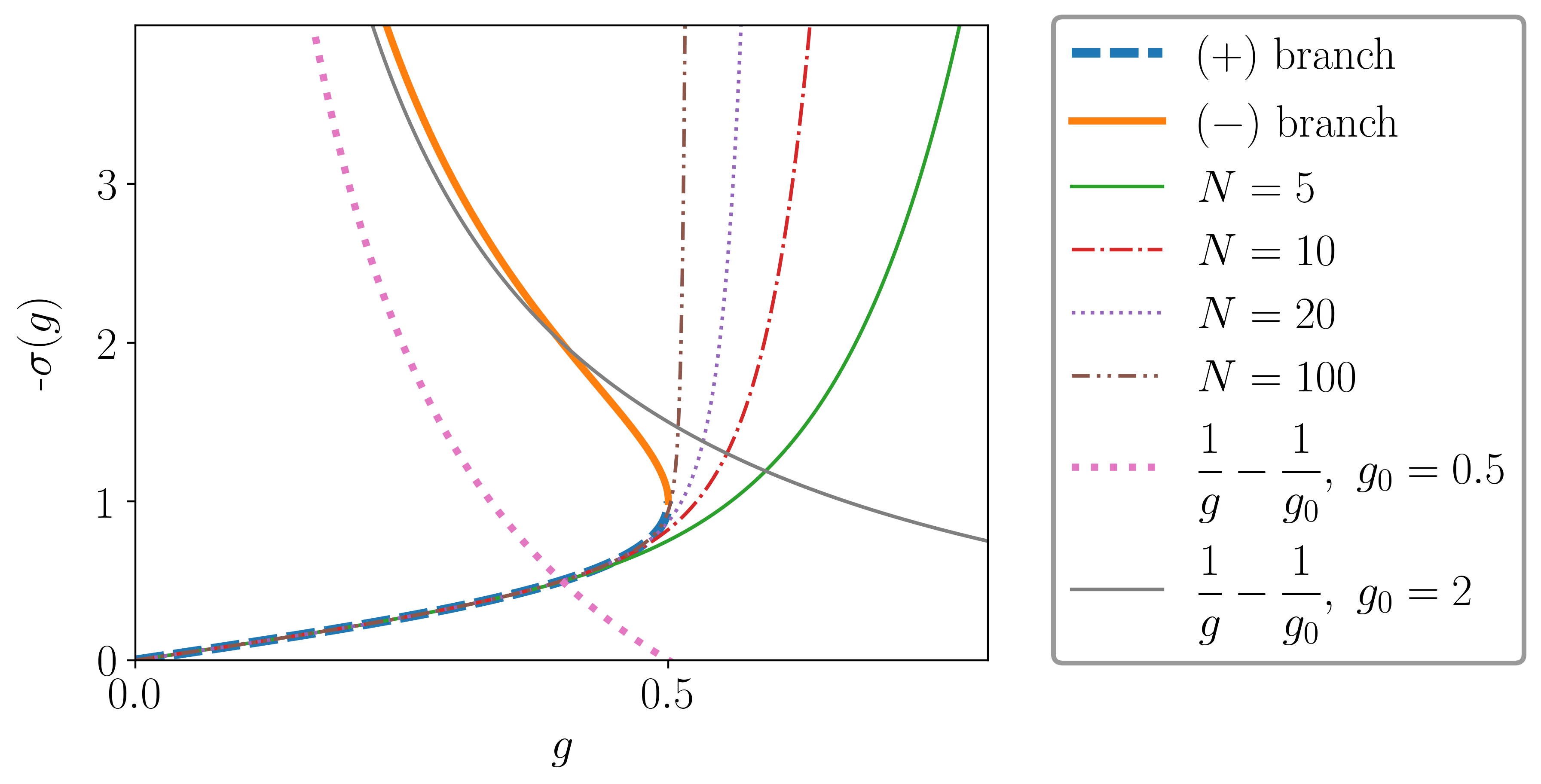}
\vskip -0.3cm
       \caption{{\it Explanation for the misleading convergence.}
         The two branches of the self-energy $\sigma^{(\pm)}(g)$, 
         together with the partial sums of the skeleton series $\sigma_{\rm bold}^{(\leq\Nr)}(g)$ for different values of the truncation order $\Nr$.
         Also shown is the curve corresponding to the Dyson equation, $-\sigma = 1/g - 1/g_0$. 
This Dyson-equation curve
         intersects
         $\sigma_{\rm bold}^{(\leq\Nr)}(g)$ at $g=g_\Nr$\,,
whereas
           the exact propagator $g=g_{\rm exact}$ is given by
           the intersection of the Dyson-equation curve with the
physical branch
           $\sigma^{({\rm sign}(1-g_0))}(g)$\,.
         It appears clearly that for $g_0<1$, $g_\Nr$ converges to the exact $g$, while for $g_0>1$, $g_\Nr$ always tends to $1/2$, the convergence radius of the skeleton series.}
       \label{fig:sigma_gN}
    \end{center}
    \vskip -0.3cm
       \end{figure}

Let us focus on the regime $g_0>1$,
where the convergence to an unphysical result takes place
(as demonstrated in Fig.~\ref{fig:cvg_wrong}).
The fact that the skeleton sequence converges at all in this regime
     is non-trivial. 
     The value of the unphysical limit $g_\infty=1/2$ of the skeleton sequence $g_\Nr$
     is equal to
     the radius of convergence of the skeleton series $\sigma_{\rm bold}(g)$.
     {This is not a coincidence, and}
     the reason for this self-tuning towards the convergence radius
     becomes clear from Fig.~\ref{fig:sigma_gN}:
     {For a large truncation order, the curve representing the truncated skeleton series as a function of $g$ becomes an almost vertical line above the position of the convergence radius ($g=1/2$), so that it intersects the Dyson-equation curve near this value of $g$.}
It also becomes clear that we are in an unusual situation where
\be
\lim_{\Nr\to\infty} \sigma_{\rm bold}^{(\leq\Nr)}(g_\Nr) \ \neq\  \lim_{\Nr\to\infty} \sigma_{\rm bold}^{(\leq\Nr)}(g_{\infty})
\ \equiv \
\sigma_{\rm bold}(g_\infty)\,.
\label{eq:lim_non_comm}
\ee



          \subsection{Diagnosing the misleading convergence}

          In the general case where $\Gex$ is unknown,
when
one observes numerically that $\GtN$ converges to some limit,
one needs a way to
tell whether this limit is equal to $\Gex$,
{\it i.e.},
whether the result can be trusted.
To this end, we consider
\be
\Sigma_{\Nr, \xi} \ \coloneqq \ \sum_{n=1}^{N} \ \Sigma_{\rm bold}^{(n)}[\GtN]\ \ \xi^n.
\label{eq:def_sig_xi}
\ee
Assuming that
         $\GtN \to \Gtinf$ for $N\to\infty$,
the following criterion~\cite{ShiftedAction} is a sufficient condition
for $\Gtinf$ to be equal to 
$\Gex$\,:
\be
\left\{
\begin{tabular}{l}
  {\it There exists $\epsilon>0$ such that:}
  \\ {\it For any $\xi$ in the disc $\Dr=\{\,|\xi|< 1+\epsilon \,\}$\,,} 
{\it \ $\Sigma_{N,\xi}$ converges for $\Nr\to\infty$;}
\\ {\it moreover, this sequence is  uniformly bounded for $\xi \in \Dr$.}
\end{tabular}
\right\}
\label{eq:criter}
\ee 
The derivation of this criterion is contained in~\cite{ShiftedAction}, and its main steps are reproduced in the Appendix for convenience.

For all practical purpose, we expect
the criterion (\ref{eq:criter}) to be essentially equivalent to the following simpler one:
\begin{equation}
  \textrm{\it There exists {$\xi>1$} 
    such that}
  \ \ \Sigma_{N, {\xi}}\
\textrm{\it converges for } \Nr\to\infty.
\label{eq:manifesto_crit}
\end{equation}
{\bl Indeed, (\ref{eq:criter}) implies (\ref{eq:manifesto_crit}), 
  and a situation where (\ref{eq:manifesto_crit}) would hold while (\ref{eq:criter}) would not hold seems unlikely to occur.} 
In what follows we will use the simplified criterion (\ref{eq:manifesto_crit}).
  We also introduce an extra factor $1/\xi^{N_0}$ in the definition (\ref{eq:def_sig_xi}) of $\Sigma_{N,\xi}$\,, where 
  {the value of $N_0$ will be conveniently chosen;
    such an $N$-independent factor does not matter for the criterion
    (it does not change whether or not the sequence $\Sigma_{N,\xi}$ converges).}

\begin{figure*}[h!]
  \vskip 0.3cm
\begin{minipage}[l]{0.49\linewidth}
       \includegraphics[width=\linewidth,trim=7 0 10 0, clip]{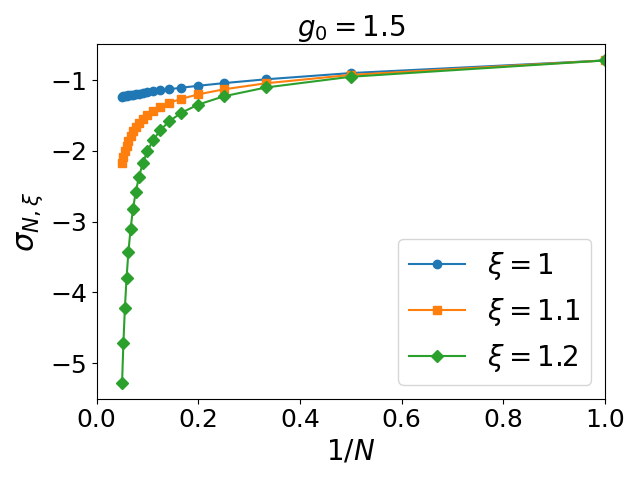}
\end{minipage}
\hskip 0.01\linewidth
\begin{minipage}[r]{0.49\linewidth}
       \includegraphics[width=\linewidth,trim=7 0 10 0, clip]{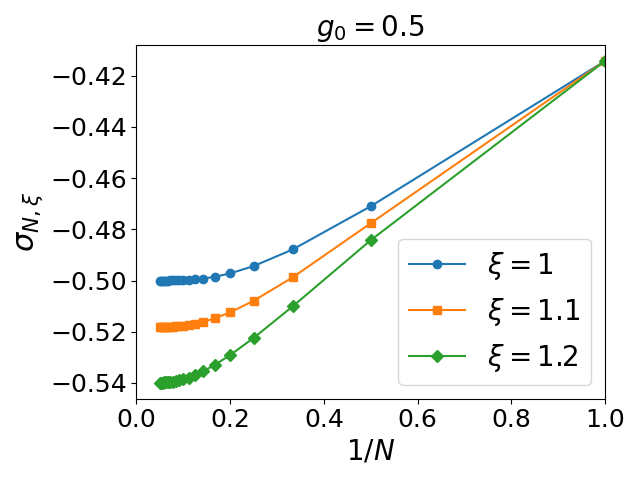}
\end{minipage}
\caption{{\it Detecting the misleading convergence for the toy model.}
  Introducing a finite~$\xi$, the sequence becomes divergent which enables one to detect the problem (left panel), or remains convergent which enables one to trust the result (right panel).
       \label{fig:thm_manif}
}
\vskip -0.3cm
       \end{figure*}

For the toy-model, this means
that assuming $g_\Nr \to g_\infty$ for $\Nr\to\infty$,
a sufficient condition for $g_\infty$ to be equal to the correct physical $g_{\rm exact}(g_0)$ is that
{there exists $\xi>1$ such that}
\be
\sigma_{N, \xi}
\ \coloneqq \,
\ \sum_{n=1}^\Nr \ \sigma_{\rm bold}^{(n)}(g_\Nr)\
{\xi}^{n-1}
\ = \ \sigma_{\rm bold}^{(\leq \Nr)}(g_\Nr\sqrt{\xi}\,)\ /\ \sqrt{\xi}
\non 
\ee
converges for $\Nr\to\infty$.
As illustrated in Fig.~\ref{fig:thm_manif},
this criterion indeed enables one to detect the misleading convergence for $g_0>1$,
and to trust the result for $g_0<1$.\footnote{For the toy model, the criterion is easily understood from Fig.~\ref{fig:sigma_gN}. For $g_0>1$,
  $g_\Nr\to1/2$ for $\Nr\to\infty$,
so that for any $\xi>1$, $\lim_{\Nr\to\infty}g_\Nr\sqrt{\xi}$ is strictly larger than the convergence radius $1/2$, leading to the divergence of $\sigma_{\rm bold}^{(\leq \Nr)}(g_\Nr\sqrt{\xi}\,)$.
On the contrary, for $g_0<1$, the $g_\Nr$'s stay at a finite distance on the left of the convergence radius $1/2$.
}


\section{Hubbard atom}

We turn to the single-site Hubbard model,
defined by the grand-canonical Hamiltonian
$-\mu\sum_\spin n_\spin$ $+$ $U\, n_\up n_\down$.
The propagator can be expressed as a functional integral over
 $\beta$-antiperiodic Grassmann fields~\cite{NegeleOrland,DupuisVol1},
\be
G_\spin(\tau) \ = \
-\ \la \, \varphi_\spin(\tau)\,\overbar{\varphi}_\spin(0) \, \ra_S
\  \equiv \ - \ \frac{\int \Dr \varphi \, \Dr \overbar{\varphi}\ \ \varphi_\spin(\tau)\,\overbar{\varphi}_\spin(0)\ e^{-S}}{\int \Dr \varphi \, \Dr \overbar{\varphi}\ e^{-S}}
\label{eq:G_hubbat}
\ee
with the action
\be
S = \int_0^\beta d\tau\  \left[
  -\sum_\spin \overbar{\varphi}_\spin(\tau) (G_0^{-1}\,\varphi_\spin)(\tau)
  \, + \, U\ (\overbar{\varphi}_\up \overbar{\varphi}_\down \varphi_\down \varphi_\up)(\tau)
  \right]
\label{eq:S_hubbat}
\ee
and
\be
G_0^{-1} \ = \ \mu \ - \ \frac{d}{d\tau}\,.
\label{eq:G0_hubbat}
\ee

We restrict for simplicity to the half-filled case
$\mu = U/2$\,, which should be the most dangerous case, since it is at and around half-filling that  the misleading convergence of $\Sigma_{\rm bold}[G_{\rm exact}]$ was discovered in~\cite{KFG}.
We use the BDMC method~\cite{ProkofevSvistunovPolaronLong,VanHoucke1,VanHouckeEOS,BDMC_long} to
sum all skeleton diagrams and
  solve the self-consistency equation (\ref{eq:bold_N}) for truncation orders $N\leq8$ {(note that at half filling, $\Sigma_{\rm bold}^{(n)} = 0$ for all odd $n>1$).}

\begin{figure*}[h!]
\begin{minipage}[l]{0.49\linewidth}
       \includegraphics[width=\linewidth,trim=60 50 120 50, clip]{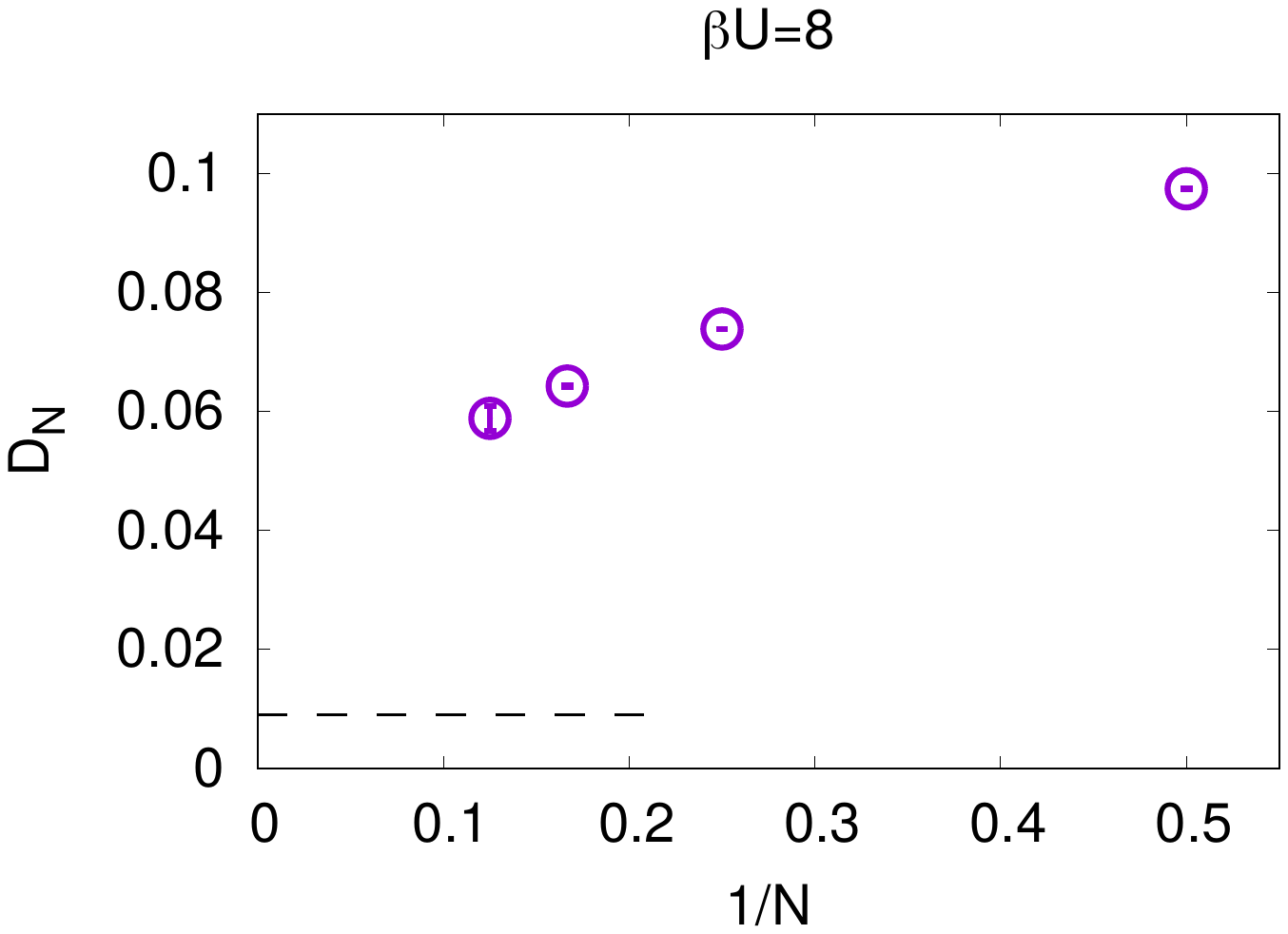}
\end{minipage}
\hskip 0.01\linewidth
\begin{minipage}[r]{0.49\linewidth}
       \includegraphics[width=\linewidth,trim=60 50 120 50, clip]{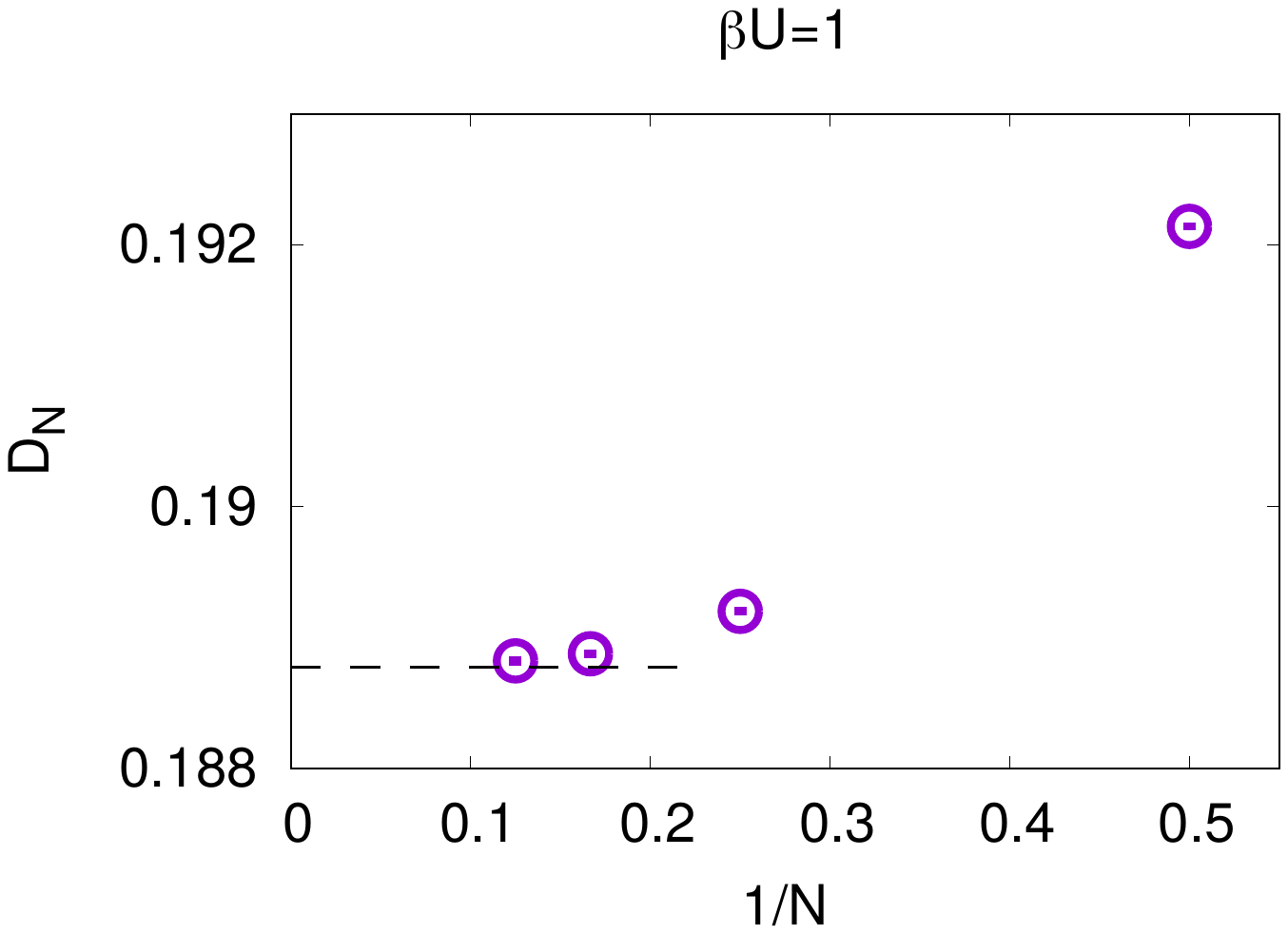}
\end{minipage}
       \caption{For the Hubbard atom at half filling, the double occupancy, as obtained from the skeleton sequence, converges to an unphysical result for large $U$ (left panel) and to the correct result for small enough $U$ (right panel) when the truncation order $\Nr\to\infty$ (dashed line: exact result).
       \label{fig:hubbat_D}
}
       \end{figure*}

The first question is whether the skeleton sequence $\GtN$ can also converge to an unphysical result,
or equivalently, whether
$\Sigma_{\rm bold}^{(\leq \Nr)}[\GtN] =: \tSig_\Nr$
can converge to an unphysical result.
Let us first consider the double occupancy
\be
D = \la n_\up n_\down \ra = U^{-1}\,{\rm tr}\,(\Sigma\,G)
\non
\ee
and the corresponding sequence $\tD_\Nr := U^{-1}\,{\rm tr}\,(\tSig_\Nr\,\GtN)$.
At large enough $U$, our data strongly indicate that this sequence does converge (albeit slowly) towards an unphysical result,
see left panel of Fig.~\ref{fig:hubbat_D}.
For small enough $U$, there is a fast convergence to the correct result,
see right panel of Fig.~\ref{fig:hubbat_D}.

\begin{figure*}[h!]
            \begin{center}
  \includegraphics[width=0.57\linewidth,trim=50 10 110 50, clip]{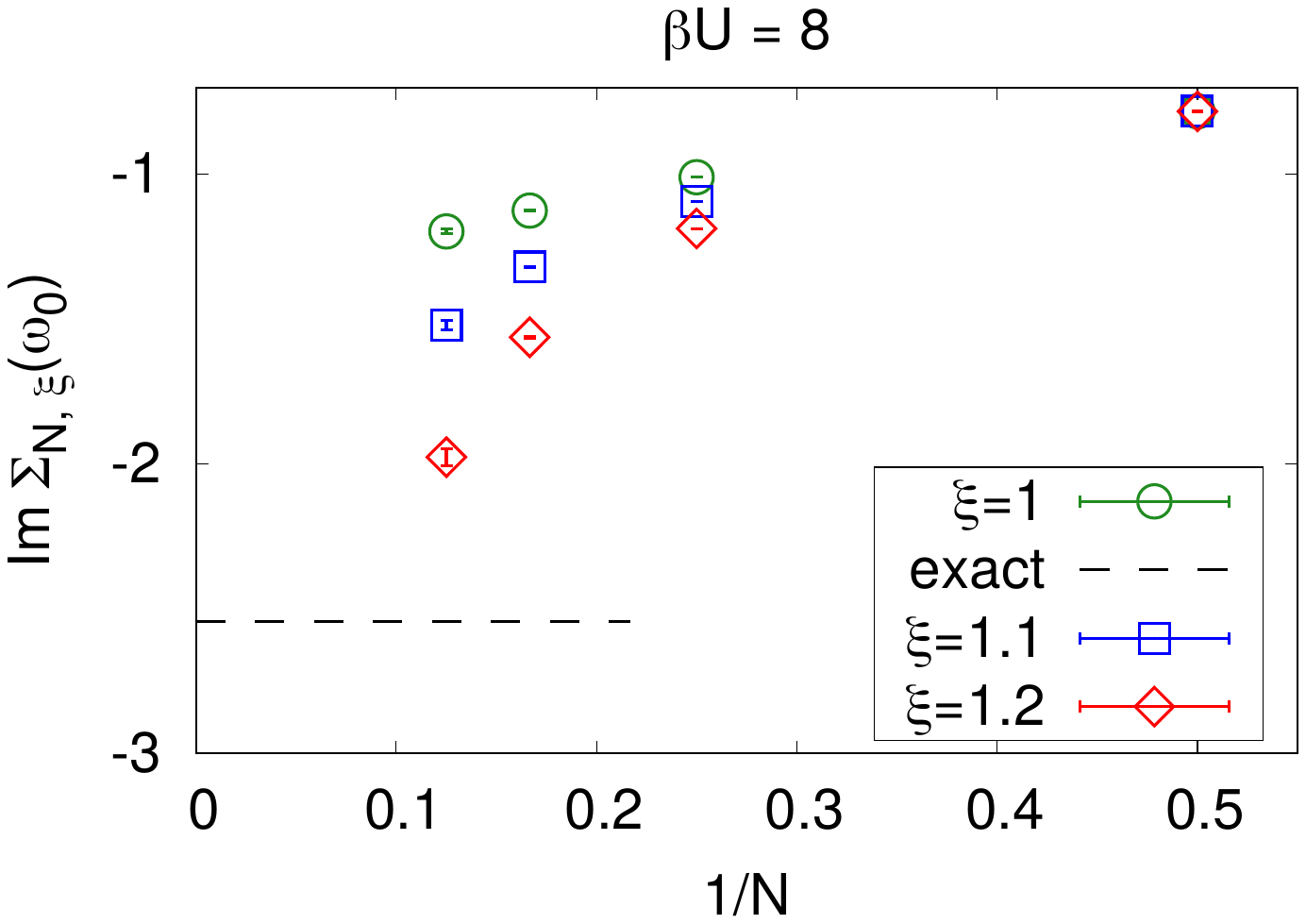}
    \vskip -0.6cm
  \caption{For the half-filled Hubbard atom at large coupling, the original skeleton sequence
    $(\xi =1)$
    converges to an unphysical result.
    At
${\xi > 1}$,
    the sequence does not converge any more: The data do not satisfy the criterion.
  }
    \label{fig:hubbat_criter_bU8}
  \vskip -0.2cm
  
  \includegraphics[width=0.57\linewidth,trim=50 -20 110 0, clip]{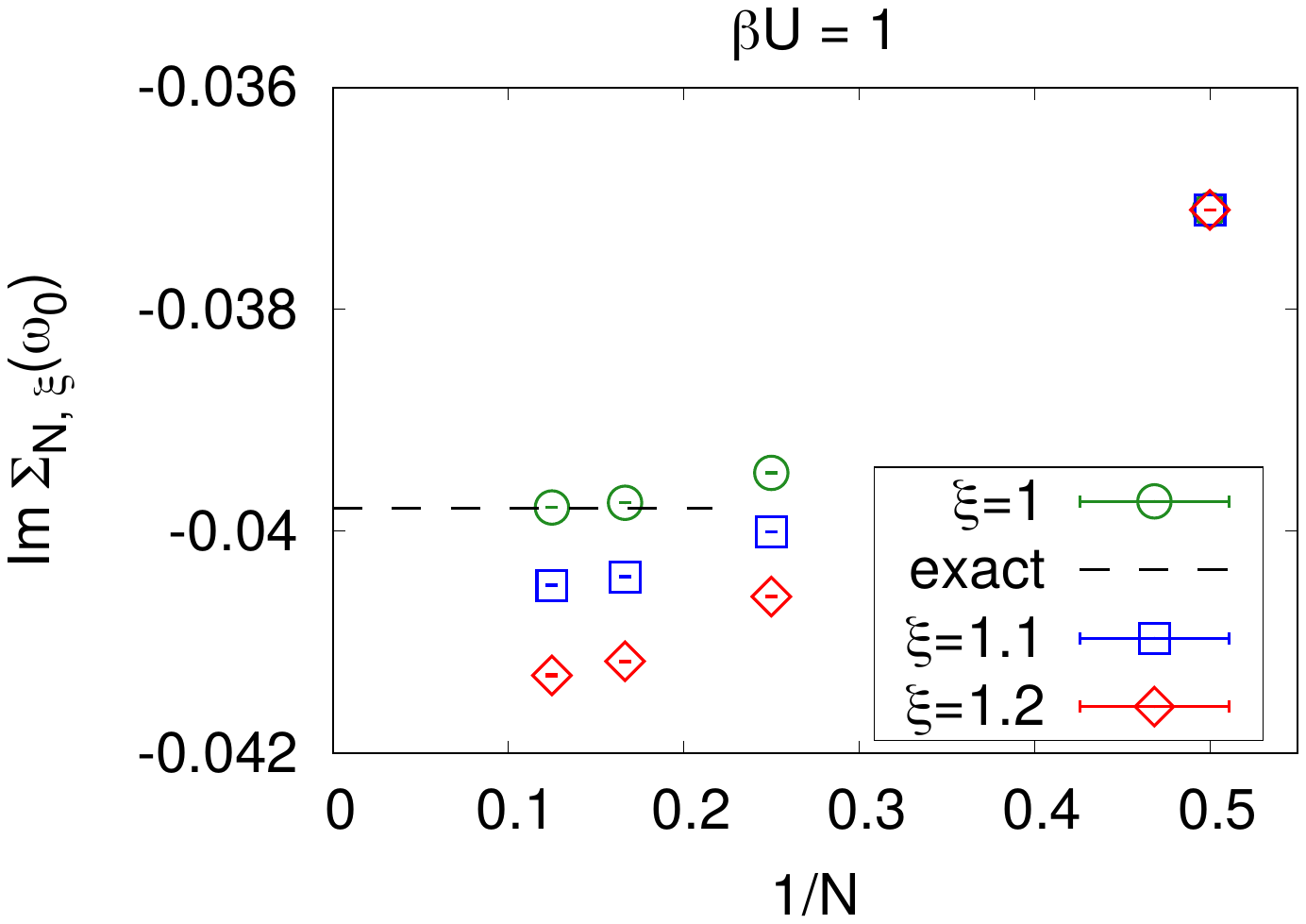}
    \vskip -0.9cm
    \caption{For the half-filled Hubbard atom at
      small enough coupling,
      the original skeleton sequence
         ({$\xi = 1$})
         converges to the correct physical result.
         At
{$\xi>1$},
the sequence remains convergent:
The criterion enables one to trust the result.
       \label{fig:hubbat_criter_bU1}
}
       \end{center}
       \end{figure*}

          The next question is whether the criterion~(\ref{eq:manifesto_crit})
          enables us to discriminate between these two situations.
We therefore plot
the sequence $\Sigma_{N,{\xi}
}$
in Figs.~\ref{fig:hubbat_criter_bU8} and~\ref{fig:hubbat_criter_bU1}.
We only show the imaginary part
because in
the considered half-filled case, 
the real part
of
{$\Sigma_{N}(\omega_n)$
automatically}
equals $U/2$;
moreover we focus for simplicity on the lowest Matsubara frequency $\omega_0 = \pi/\beta$,
{and we choose $N_0=2$.}

For ${\xi=1}$, 
$\Sigma_{N,{\xi}}$ 
reduces to the original skeleton sequence $\tSig_\Nr$,
and the behavior is similar to the double occupancy:
The sequence appears to
converge, albeit slowly, towards an unphysical result for $\beta U=8$ (Fig.~\ref{fig:hubbat_criter_bU8}), while fast convergence to the correct physical result takes place for $\beta U=1$ (Fig.~\ref{fig:hubbat_criter_bU1}).
For
${\xi>1}$,
the sequence does not appear to converge any more 
for $\beta U=8$, see Fig.~\ref{fig:hubbat_criter_bU8}:
The data do not satisfy the criterion, indicating
that the results cannot be trusted in this case.
In contrast, for $\beta U=1$,
the criterion enables one to validate the results,
since the sequence remains convergent at
${\xi>1}$,
see Fig.~\ref{fig:hubbat_criter_bU1}.

Regarding the choice of ${\xi}$,
 it should be neither too small in order to have an effect at the accessible orders, nor too large to avoid making the criterion too conservative.
More precisely, $\xi-1$ should not be too small,
  so that
  $\xi^N$ differs significantly from 1
  (and hence $\Sigma_{N,\xi}$ differ significantly from $\Sigma_{N,1}=\Sigma_N$)
    at the largest accessible order $N_{\rm max}$\,.
This necessity to work with a finite $\xi-1$ implies that the criterion is conservative:
It leads to discarding results
in a region of the parameter space near but outside the misleading-convergence region.\footnote{The required $\xi-1$ scales as $1/N_{\rm max}$;
for the toy model this leads to discarding results
not only in the misleading-convergence region $g_0>1$,
but also for $0\leq 1-g_0 \lesssim 1/N_{\rm max}$.}
For~$N_{\rm max}=8$,
the choices $\xi-1 = 0.1$ and 0.2
are {\it a priori} large enough
(since $\xi^8 \approx 2$ and 4)
and
Fig.~\ref{fig:hubbat_criter_bU8}
confirms that they allow to detect the divergence of $\Sigma_{N,\xi}$ in the misleading-convergence regime,
while on the other hand Fig.~\ref{fig:hubbat_criter_bU1} shows
that $\beta U=1$ is at a sufficient distance  from the misleading-convergence region
for $\Sigma_{N,\xi}$ to remain convergent for these $\xi$ values.


We remark
that
when solving Eq.~(\ref{eq:bold_N}) by iterations,
in the case where the convergence to the unphysical result for $\Nr\to\infty$ occurs,
convergence as a function of iterations at fixed $\Nr$
only takes place
if we use
a damping procedure,
where  $\GtN$ at iteration $(i+1)$
is
obtained as
$\GtN^{(i+1)} = [G_0^{-1} - \Sigma^{(i)}]^{-1}$
with $\Sigma^{(i)}$ a
weighted average of\, $\Sigma_{\rm bold}^{(\leq\Nr)}[\GtN^{(i)}]$ and $\Sigma^{(i-1)}$,
while the fixed point is unstable for the undamped iterative procedure $\Sigma^{(i)} \coloneqq \Sigma_{\rm bold}^{(\leq\Nr)}[\GtN^{(i)}]$.
Such a damping procedure is commonly used in BDMC { where it also reduces} the statistical error~\cite{ProkofevSvistunovBoldPRL,BDMC_long}.
In the toy model,
one can easily show that
an increasingly strong damping is required when $\Nr$ is increased,
because for $\Nr\to\infty$, the slope $[d\sigma_{\rm bold}^{(\leq\Nr)}(g)/dg]_{g=g_\Nr}$
diverges,
{ making the undamped iterative procedure unstable}.
This observation could also 
be useful 
for misleading-convergence detection.

Finally, we comment on
          the link with the multivaluedness of the self-energy functional $\Sigma[G]$ ({\it i.e.}, of the Luttinger-Ward functional).
          In~\cite{KFG},
          the misleading convergence of the skeleton series
          was found to be towards an unphysical branch of the self-energy functional,
          in the sense that if Eqs.~(\ref{eq:G_hubbat},\ref{eq:S_hubbat}) are viewed as a mapping $G_0 \mapsto G[G_0]$,
          then there exists $G_{0, {\rm unphys}}$ such that            $\Sigma_{\rm bold}[\Gex] = G_{0, {\rm unphys}}^{-1} - \Gex^{-1}$            and $G[G_{0, {\rm unphys}}] = \Gex \equiv G[G_0]$.
          As noted in~\cite{KFG}, this
          $G_{0, {\rm unphys}}$ does not belong to the set of physical bare propagators which are of the form (\ref{eq:G0_hubbat}) for some value of chemical potential;
          therefore, by looking at $G_{0, {\rm unphys}}$\,,
          one can tell that the result is on an unphysical branch, and hence
detect the misleading convergence of the skeleton series.
In contrast, the misleading convergence of the 
sequence $\GtN$
reported here cannot be detected in this way.
Indeed, the self-consistency equation~(\ref{eq:bold_N}) is enforced
with the original physical $G_0$.

          \section{Conclusion}
          
          We have demonstrated that there is a regime where
the solution of self-consistent many-body perturbation theory
converges to an unphysical result in the limit of infinite truncation order 
of the skeleton series.
This surprising breakdown
of the standard framework
results
  from the findings of~\cite{KFG}
  combined with an additional subtle mathematical mechanism which we have elucidated by analyzing the zero space-time dimensional model of~\cite{Rossi_Werner_0+0dim}.
          In this problematic regime, lowest order calculations
can be off by one order of magnitude,
       but access to higher orders
    enables one
to detect
the problem
numerically through the divergence of a slightly modified sequence,
whereas seeing convergence of this modified sequence enables one to rule out misleading convergence and to trust the result,
as proposed in \cite{ShiftedAction} and demonstrated here for the Hubbard atom.
Such a proof of principle is relevant for many-body problems
  in regimes where,
    in spite of 
    important progress
  with non self-consistent 
  frameworks~\cite{RossiCDet,SimkovicCDet,MoutenetCDet,SimkovicCrossover,KimKozik2020,JohanInfiniteU,KunHauleEG,HauleChenEG_2,GullPRR2020,SimkovicLadderCDet,RossiRDet,SimkovicSuscept_k,LenihanTc,SpadaSF,SimkovicPG,WietekTriang,ChenWDM,TupitsynRealFreq,LeblancTupitsynEG,TupitsynLandauDamping,KozikCombinatorialSummation,RossiNesting} (for which it was shown that misleading convergence generically does not occur~\cite{ShiftedAction}\,)
  and with strong-coupling expansions~\cite{CarlstromStrongCouplingDiagMC,CarlstromTMD,CarlstromSpectralTopol},
 self-consistent BDMC remains among the state of the art approaches.
 In particular,
during finalization of the present manuscript,
its main findings have been used as a basis
to discriminate between physical and unphysical
BDMC results for the doped two-dimensional Hubbard model at strong coupling in a non-Fermi liquid regime~\cite{KimStrangeMetal}.

\section*{Acknowledgments}

We thank N.~Prokof'ev, B.~Svistunov and L.~Reining for useful discussions and comments.

\paragraph{Funding information}

We acknowledge support from
ERC ({\it Critisup2}, H2020 Adv-743159) (F.W.),
ANR ({\it LODIS}, ANR-21-CE30-0033) (K.V.H. and F.W.),
the Simons Foundation through the Simons Collaboration on the Many Electron Problem and EPSRC (EP/P003052/1) (E.K.),
the National Natural Science Foundation of China (grant No. 11625522) and the Science and Technology Committee of Shanghai (grant No. 20DZ2210100) (Y.D.).
The Flatiron Institute is a division of the Simons Foundation.
          

\newpage
\begin{appendix}


    \subsection*{Appendix: Main steps of the derivation of the criterion}
    
      \addcontentsline{toc}{section}{Appendix} 

\setcounter{section}{1}

For convenience, we reproduce here the main steps of the derivation of the criterion~(\ref{eq:criter}), see~\cite{ShiftedAction,InPrepFFS} for more details.
For definiteness, we consider the Hubbard model at finite temperature;
the reasoning is also directly applicable to the Hubbard atom (by removing the position variable) and to the zero space-time dimensional toy model (by also removing the imaginary time variable).
Let
$\Sigma_{\infty, \xi} 
\ := \  \lim_{N\to\infty} \ \Sigma_{N, \xi}$\,. 
By making use of Morera's theorem, Cauchy's integral formula and the dominated convergence theorem, the condition~(\ref{eq:criter}) allows one to show the key property
\be
\Sigma_{\infty, \xi} 
\ = \ \sum_{n=1}^\infty \ \Sigma_{\rm bold}^{(n)}[\Gtinf]
\ \, \xi^n\,,\ \ \forall \xi\in\Dr\,.
\label{eq:sig_inf=sig_bold}
\ee
Setting $\xi=1$ in (\ref{eq:sig_inf=sig_bold}) 
yields\footnote{Equation~(\ref{eq:sig_bold_N->sig_bold}) breaks down for the toy model in the problematic regime ($g_0>1$), as pointed out in Eq.~(\ref{eq:lim_non_comm}).}
\be
\lim_{N\to\infty} \ \Sigma_{\rm bold}^{(\leq N)}[G_N] 
\ = \ \Sigma_{\rm bold}[G_\infty]
\,.
  \label{eq:sig_bold_N->sig_bold}
  \ee
  Substituting (\ref{eq:sig_bold_N->sig_bold}) into (\ref{eq:bold_N}) yields
\be
G_\infty^{-1} 
\ = \ G_0^{-1} 
\ - \ \Sigma_{\rm bold}[G_\infty]\,. 
\label{eq:Ginf_dyson}
\ee
  The next step is to consider the action
\bea
  S_{\rm bold}^{(\xi)}  &:=& 
  \ -\ \sum_{\rr, s}  \int_0^\beta d\tau \  \left[ \overbar{\varphi}_s \left( \Gtinf^{-1} \ + \ \sum_{n=1}^\infty \ \Sigma_{\rm bold}^{(n)}[\Gtinf]\ \xi^n  \right)\, \varphi_s \right]\!(\rr,\tau)
  \non
  \\ && \ + \ \xi\  U\ \sum_{\rr} \int_0^\beta d\tau \ (\overbar{\varphi}_\up \overbar{\varphi}_\down \varphi_\down \varphi_\up)(\rr,\tau)
  \non
  \eea
and the corresponding propagator
$G_{\rm bold}^{(\xi)}(\rr,\tau) \ := \ - \, \la \, \varphi_s(\rr,\tau)\ \overbar{\varphi}_s(\vn,0)\,\ra_{S_{\rm bold}^{(\xi)}}\,.$
The action $S_{\rm bold}^{(\xi)}$ is designed in such a way that
\be
\left. \frac{\partial^n G_{\rm bold}^{(\xi)}}{\partial \xi^n}\right|_{\xi=0} \ = 0\,,\ \ \forall n\geq1\,.
\label{eq:derGbold=0}
\ee
Obviously, $G_{\rm bold}^{(\xi=0)} = \Gtinf$\,.
On the other hand, (\ref{eq:Ginf_dyson}) implies that $S_{\rm bold}^{(\xi=1)}$ is equal to the physical action of the Hubbard model, so that $G_{\rm bold}^{(\xi=1)} = G_{\rm exact}$\,.
Now, since $S_{\rm bold}^{(\xi)}$ depends analytically on $\xi$ in $\Dr$, we expect (at least for fermions on a lattice at finite temperature) that one of the following alternatives holds:
\bi
\item[{\it (i)}] $G_{\rm bold}^{(\xi)}$ depends analytically on $\xi$ in $\Dr$ 
\item[{\it (ii)}] $G_{\rm bold}^{(\xi)}$ has a non-removable singularity at a point $\xi_c\in\Dr$\, ({\it e.g.}, a phase transition),
  \\and $G_{\rm bold}^{(\xi)}$ is analytic in the disc $\{\,|\xi|<|\xi_c|\,\}$\,.
  \ei
In case {\it (ii)}, the convergence radius of the Taylor series of $G_{\rm bold}^{(\xi)}$ at the origin would be \,$|\xi_c|$\,, in contradiction with (\ref{eq:derGbold=0}). Hence {\it (i)} holds, and we have
$$G_{\rm exact} \,=\  G_{\rm bold}^{(\xi=1)} \ =\ G_{\rm bold}^{(\xi=0)} \ +\ \,\sum_{n=1}^\infty \ \frac{1}{n!}\,\left. \frac{\partial^n G_{\rm bold}^{(\xi)}}{\partial \xi^n}\right|_{\xi=0} \ =\   G_{\rm bold}^{(\xi=0)} \ =\  \Gtinf\,.$$
{\bl This concludes the derivation of the equality $\Gtinf = \,G_{\rm exact}$ under the assumption (\ref{eq:criter})\,.}

\end{appendix}

\let\oldaddcontentsline\addcontentsline
\renewcommand{\addcontentsline}[3]{}
\bibliography{felix_copy}

\begin{thebibliography}{10}
\providecommand{\url}[1]{\texttt{#1}}
\providecommand{\urlprefix}{URL }
\expandafter\ifx\csname urlstyle\endcsname\relax
  \providecommand{\doi}[1]{doi:\discretionary{}{}{}#1}\else
  \providecommand{\doi}{doi:\discretionary{}{}{}\begingroup
  \urlstyle{rm}\Url}\fi
\providecommand{\eprint}[2][]{\url{#2}}

\bibitem{KFG}
E.~Kozik, M.~Ferrero and A.~Georges,
\newblock \emph{Nonexistence of the Luttinger-Ward Functional and Misleading
  Convergence of Skeleton Diagrammatic Series for Hubbard-Like Models},
\newblock Phys. Rev. Lett. \textbf{114}, 156402 (2015),
\newblock \doi{10.1103/PhysRevLett.114.156402}.

\bibitem{ShiftedAction}
R.~Rossi, F.~Werner, N.~Prokof'ev and B.~Svistunov,
\newblock \emph{Shifted-Action Expansion and Applicability of Dressed
  Diagrammatic Schemes},
\newblock Phys. Rev. B \textbf{93}, 161102(R) (2016),
\newblock \doi{10.1103/PhysRevB.93.161102}.

\bibitem{MartinReiningCeperley}
R.~M. Martin, L.~Reining and D.~M. Ceperley,
\newblock \emph{Interacting Electrons: Theory and Computational Approaches},
\newblock Cambridge University Press,
\newblock ISBN 9781139050807,
\newblock \doi{10.1017/CBO9781139050807} (2016).

\bibitem{DupuisVol1}
N.~Dupuis,
\newblock \emph{Field Theory of Condensed Matter and Ultracold Gases},
\newblock World Scientific,
\newblock ISBN 978-1-80061-390-4,
\newblock \doi{10.1142/q0409} (2023).

\bibitem{StefanucciVanLeeuwen}
G.~Stefanucci and R.~van Leeuwen,
\newblock \emph{Nonequilibrium Many-Body Theory of Quantum Systems: A Modern
  Introduction},
\newblock Cambridge University Press,
\newblock ISBN 9781139023979,
\newblock \doi{10.1017/CBO9781139023979} (2013).

\bibitem{VanHoucke1}
K.~{Van Houcke}, E.~Kozik, N.~Prokof'ev and B.~Svistunov,
\newblock {\it Diagrammatic Monte Carlo}, in Computer Simulation Studies in
  Condensed Matter Physics XXI. CSP-2008. Eds. D.P. Landau, S.P. Lewis, and
  H.B. Sch{\"u}ttler, Phys. Procedia {\bf 6}, 95 (2010),
\newblock \doi{10.1016/j.phpro.2010.09.034}.

\bibitem{reining_2solutions}
A.~Stan, P.~Romaniello, S.~Rigamonti, L.~Reining and J.~Berger,
\newblock \emph{Unphysical and physical solutions in many-body theories: from
  weak to strong correlation},
\newblock New J. Phys. \textbf{17}, 093045 (2015),
\newblock \doi{10.1088/1367-2630/17/9/093045}.

\bibitem{Rossi_Werner_0+0dim}
R.~Rossi and F.~Werner,
\newblock \emph{Skeleton series and multivaluedness of the self-energy
  functional in zero space-time dimensions},
\newblock J. Phys. A \textbf{48}, 485202 (2015),
\newblock \doi{10.1088/1751-8113/48/48/485202}.

\bibitem{SchaeferDivergences2016}
T.~Sch\"afer, S.~Ciuchi, M.~Wallerberger, P.~Thunstr\"om, O.~Gunnarsson,
  G.~Sangiovanni, G.~Rohringer and A.~Toschi,
\newblock \emph{Nonperturbative landscape of the Mott-Hubbard transition:
  Multiple divergence lines around the critical endpoint},
\newblock Phys. Rev. B \textbf{94}, 235108 (2016),
\newblock \doi{10.1103/PhysRevB.94.235108}.

\bibitem{ReiningHubbat}
W.~Tarantino, P.~Romaniello, J.~A. Berger and L.~Reining,
\newblock \emph{Self-consistent Dyson equation and self-energy functionals: An
  analysis and illustration on the example of the Hubbard atom},
\newblock Phys. Rev. B \textbf{96}, 045124 (2017),
\newblock \doi{10.1103/PhysRevB.96.045124}.

\bibitem{GunnarsonBreakdownPRL}
O.~Gunnarsson, G.~Rohringer, T.~Sch\"afer, G.~Sangiovanni and A.~Toschi,
\newblock \emph{Breakdown of Traditional Many-Body Theories for Correlated
  Electrons},
\newblock Phys. Rev. Lett. \textbf{119}, 056402 (2017),
\newblock \doi{10.1103/PhysRevLett.119.056402}.

\bibitem{ParcolletMultival}
J.~Vu\ifmmode \check{c}\else \v{c}\fi{}i\ifmmode \check{c}\else
  \v{c}\fi{}evi\ifmmode~\acute{c}\else \'{c}\fi{}, N.~Wentzell, M.~Ferrero and
  O.~Parcollet,
\newblock \emph{Practical consequences of the Luttinger-Ward functional
  multivaluedness for cluster DMFT methods},
\newblock Phys. Rev. B \textbf{97}, 125141 (2018),
\newblock \doi{10.1103/PhysRevB.97.125141}.

\bibitem{LinLin}
L.~Lin and M.~Lindsey,
\newblock \emph{Variational structure of Luttinger{\textendash}Ward formalism
  and bold diagrammatic expansion for Euclidean lattice field theory},
\newblock PNAS \textbf{115}(10), 2282 (2018),
\newblock \doi{10.1073/pnas.1720782115}.

\bibitem{ToschiDivergences2018}
P.~Chalupa, P.~Gunacker, T.~Sch\"afer, K.~Held and A.~Toschi,
\newblock \emph{Divergences of the irreducible vertex functions in correlated
  metallic systems: Insights from the Anderson impurity model},
\newblock Phys. Rev. B \textbf{97}, 245136 (2018),
\newblock \doi{10.1103/PhysRevB.97.245136}.

\bibitem{KimMultival}
A.~J. Kim and V.~Sacksteder,
\newblock \emph{Multivaluedness of the Luttinger-Ward functional in the
  fermionic and bosonic system with replicas},
\newblock Phys. Rev. B \textbf{101}, 115146 (2020),
\newblock \doi{10.1103/PhysRevB.101.115146}.

\bibitem{SchaeferDivergentPrecursors}
T.~Sch\"afer, G.~Rohringer, O.~Gunnarsson, S.~Ciuchi, G.~Sangiovanni and
  A.~Toschi,
\newblock \emph{Divergent Precursors of the Mott-Hubbard Transition at the
  Two-Particle Level},
\newblock Phys. Rev. Lett. \textbf{110}, 246405 (2013),
\newblock \doi{10.1103/PhysRevLett.110.246405}.

\bibitem{GunnarsonParquetDCA2016}
O.~Gunnarsson, T.~Sch\"afer, J.~P.~F. LeBlanc, J.~Merino, G.~Sangiovanni,
  G.~Rohringer and A.~Toschi,
\newblock \emph{Parquet decomposition calculations of the electronic
  self-energy},
\newblock Phys. Rev. B \textbf{93}, 245102 (2016),
\newblock \doi{10.1103/PhysRevB.93.245102}.

\bibitem{RohringerRevue}
G.~Rohringer, H.~Hafermann, A.~Toschi, A.~A. Katanin, A.~E. Antipov, M.~I.
  Katsnelson, A.~I. Lichtenstein, A.~N. Rubtsov and K.~Held,
\newblock \emph{Diagrammatic routes to nonlocal correlations beyond dynamical
  mean field theory},
\newblock Rev. Mod. Phys. \textbf{90}, 025003 (2018),
\newblock \doi{10.1103/RevModPhys.90.025003}.

\bibitem{ToschiDvg2020}
M.~Reitner, P.~Chalupa, L.~Del~Re, D.~Springer, S.~Ciuchi, G.~Sangiovanni and
  A.~Toschi,
\newblock \emph{Attractive Effect of a Strong Electronic Repulsion: The Physics
  of Vertex Divergences},
\newblock Phys. Rev. Lett. \textbf{125}, 196403 (2020),
\newblock \doi{10.1103/PhysRevLett.125.196403}.

\bibitem{AdlerNonperturbative}
S.~Adler, F.~Krien, P.~Chalupa-Gantner, G.~Sangiovanni and A.~Toschi,
\newblock \emph{{Non-perturbative intertwining between spin and charge
  correlations: A ``smoking gun'' single-boson-exchange result}},
\newblock SciPost Phys. \textbf{16}, 054 (2024),
\newblock \doi{10.21468/SciPostPhys.16.2.054}.

\bibitem{CarlstromLineNodeSemimetal}
C.~Naya, T.~Bertolini and J.~Carlstr\"om,
\newblock \emph{Stability of line-node semimetals with strong Coulomb
  interactions and properties of the symmetry-broken state},
\newblock Phys. Rev. Res. \textbf{5}, 013069 (2023),
\newblock \doi{10.1103/PhysRevResearch.5.013069}.

\bibitem{Gunnarson_GW_Revue}
F.~Aryasetiawan and O.~Gunnarsson,
\newblock \emph{The GW method},
\newblock Rep. Prog. Phys. \textbf{61}(3), 237 (1998),
\newblock \doi{10.1088/0034-4885/61/3/002}.

\bibitem{SimonsMolecules}
K.~T. Williams, Y.~Yao, J.~Li, L.~Chen, H.~Shi, M.~Motta, C.~Niu, U.~Ray,
  S.~Guo, R.~J. Anderson, J.~Li, L.~N. Tran \emph{et~al.},
\newblock \emph{Direct Comparison of Many-Body Methods for Realistic Electronic
  Hamiltonians},
\newblock Phys. Rev. X \textbf{10}, 011041 (2020),
\newblock \doi{10.1103/PhysRevX.10.011041}.

\bibitem{Leeuwen_GW2_mol}
N.~E. Dahlen and R.~van Leeuwen,
\newblock \emph{Self-consistent solution of the Dyson equation for atoms and
  molecules within a conserving approximation},
\newblock J. Chem. Phys. \textbf{122}(16), 164102 (2005),
\newblock \doi{10.1063/1.1884965}.

\bibitem{Zgid_GW2}
J.~J. Phillips and D.~Zgid,
\newblock \emph{The description of strong correlation within self-consistent
  Green's function second-order perturbation theory},
\newblock J. Chem. Phys. \textbf{140}(24), 241101 (2014),
\newblock \doi{10.1063/1.4884951}.

\bibitem{IgorCoulombPhonon}
I.~S. Tupitsyn, A.~S. Mishchenko, N.~Nagaosa and N.~Prokof'ev,
\newblock \emph{Coulomb and electron-phonon interactions in metals},
\newblock Phys. Rev. B \textbf{94}, 155145 (2016),
\newblock \doi{10.1103/PhysRevB.94.155145}.

\bibitem{IgorDirac}
I.~S. Tupitsyn and N.~V. Prokof'ev,
\newblock \emph{Stability of Dirac Liquids with Strong Coulomb Interaction},
\newblock Phys. Rev. Lett. \textbf{118}, 026403 (2017),
\newblock \doi{10.1103/PhysRevLett.118.026403}.

\bibitem{IgorHaldane}
I.~S. Tupitsyn and N.~V. Prokof'ev,
\newblock \emph{Phase diagram topology of the Haldane-Hubbard-Coulomb model},
\newblock Phys. Rev. B \textbf{99}, 121113(R) (2019),
\newblock \doi{10.1103/PhysRevB.99.121113}.

\bibitem{SimonsHydrogenChain}
M.~Motta, D.~M. Ceperley, G.~K.-L. Chan, J.~A. Gomez, E.~Gull, S.~Guo, C.~A.
  Jim\'enez-Hoyos, T.~N. Lan, J.~Li, F.~Ma, A.~J. Millis, N.~V. Prokof'ev
  \emph{et~al.},
\newblock \emph{Towards the Solution of the Many-Electron Problem in Real
  Materials: Equation of State of the Hydrogen Chain with State-of-the-Art
  Many-Body Methods},
\newblock Phys. Rev. X \textbf{7}, 031059 (2017),
\newblock \doi{10.1103/PhysRevX.7.031059}.

\bibitem{Haussmann_Z_Phys}
R.~Haussmann,
\newblock \emph{Crossover from BCS superconductivity to Bose-Einstein
  condensation: a self-consistent theory},
\newblock Z. Phys. B \textbf{91}, 291 (1993),
\newblock \doi{10.1007/BF01344058}.

\bibitem{Haussmann_PRB}
R.~Haussmann,
\newblock \emph{Properties of a Fermi liquid at the superfluid transition in
  the crossover region between BCS superconductivity and Bose-Einstein
  condensation},
\newblock Phys. Rev. B \textbf{49}, 12975 (1994),
\newblock \doi{10.1103/PhysRevB.49.12975}.

\bibitem{HaussmannZwergerThermo}
R.~Haussmann, W.~Rantner, S.~Cerrito and W.~Zwerger,
\newblock \emph{Thermodynamics of the BCS-BEC crossover},
\newblock Phys. Rev. A \textbf{75}, 023610 (2007),
\newblock \doi{10.1103/PhysRevA.75.023610}.

\bibitem{DengEmergentBCS}
Y.~Deng, E.~Kozik, N.~V. Prokof'ev and B.~V. Svistunov,
\newblock \emph{Emergent {BCS} regime of the two-dimensional fermionic
  {Hubbard} model: Ground-state phase diagram},
\newblock Europhys. Lett. \textbf{110}, 57001 (2015),
\newblock \doi{10.1209/0295-5075/110/57001}.

\bibitem{SimkovicEmergent}
F.~\ifmmode~\check{S}\else \v{S}\fi{}imkovic, Y.~Deng and E.~Kozik,
\newblock \emph{Superfluid ground state phase diagram of the two-dimensional
  Hubbard model in the emergent Bardeen-Cooper-Schrieffer regime},
\newblock Phys. Rev. B \textbf{104}, L020507 (2021),
\newblock \doi{10.1103/PhysRevB.104.L020507}.

\bibitem{VanHouckeEOS}
K.~{Van Houcke}, F.~Werner, E.~Kozik, N.~Prokof'ev, B.~Svistunov, M.~J.~H. Ku,
  A.~T. Sommer, L.~W. Cheuk, A.~Schirotzek and M.~W. Zwierlein,
\newblock \emph{Feynman diagrams versus Fermi-gas Feynman emulator},
\newblock Nature Phys. \textbf{8}, 366 (2012),
\newblock \doi{10.1038/NPHYS2273}.

\bibitem{RossiEOS}
R.~Rossi, T.~Ohgoe, K.~{Van Houcke} and F.~Werner,
\newblock \emph{Resummation of diagrammatic series with zero convergence radius
  for strongly correlated fermions},
\newblock Phys. Rev. Lett \textbf{121}, 130405 (2018),
\newblock \doi{10.1103/PhysRevLett.121.130405}.

\bibitem{RossiContact}
R.~Rossi, T.~Ohgoe, E.~Kozik, N.~Prokof'ev, B.~Svistunov, K.~{Van Houcke} and
  F.~Werner,
\newblock \emph{Contact and momentum distribution of the unitary Fermi gas},
\newblock Phys. Rev. Lett \textbf{121}, 130406 (2018),
\newblock \doi{10.1103/PhysRevLett.121.130406}.

\bibitem{MishchenkoProkofevPRL2014}
A.~S. Mishchenko, N.~Nagaosa and N.~Prokof'ev,
\newblock \emph{Diagrammatic Monte~Carlo Method for Many-Polaron Problems},
\newblock Phys. Rev. Lett. \textbf{113}, 166402 (2014),
\newblock \doi{10.1103/PhysRevLett.113.166402}.

\bibitem{KulaginPRL}
S.~A. Kulagin, N.~Prokof'ev, O.~A. Starykh, B.~Svistunov and C.~N. Varney,
\newblock \emph{Bold Diagrammatic Monte Carlo Method Applied to Fermionized
  Frustrated Spins},
\newblock Phys. Rev. Lett. \textbf{110}, 070601 (2013),
\newblock \doi{10.1103/PhysRevLett.110.070601}.

\bibitem{HuangPyro}
Y.~Huang, K.~Chen, Y.~Deng, N.~Prokof'ev and B.~Svistunov,
\newblock \emph{{Spin-Ice State of the Quantum Heisenberg Antiferromagnet on
  the Pyrochlore Lattice}},
\newblock Phys. Rev. Lett. \textbf{116}, 177203 (2016),
\newblock \doi{10.1103/PhysRevLett.116.177203}.

\bibitem{WangCaiSpins}
T.~Wang, X.~Cai, K.~Chen, N.~V. Prokof'ev and B.~V. Svistunov,
\newblock \emph{Quantum-to-classical correspondence in two-dimensional
  Heisenberg models},
\newblock Phys. Rev. B \textbf{101}, 035132 (2020),
\newblock \doi{10.1103/PhysRevB.101.035132}.

\bibitem{AyralFunctional}
T.~Ayral and O.~Parcollet,
\newblock \emph{Mott physics and spin fluctuations: A functional viewpoint},
\newblock Phys. Rev. B \textbf{93}, 235124 (2016),
\newblock \doi{10.1103/PhysRevB.93.235124}.

\bibitem{NegeleOrland}
J.~W. Negele and H.~Orland,
\newblock \emph{Quantum Many-particle Systems},
\newblock Addison-Wesley,
\newblock \doi{10.1201/9780429497926} (1988).

\bibitem{ProkofevSvistunovPolaronLong}
N.~V. Prokof'ev and B.~V. Svistunov,
\newblock \emph{Bold diagrammatic Monte Carlo: A generic sign-problem tolerant
  technique for polaron models and possibly interacting many-body problems},
\newblock Phys. Rev. B \textbf{77}, 125101 (2008),
\newblock \doi{10.1103/PhysRevB.77.125101}.

\bibitem{BDMC_long}
K.~{Van Houcke}, F.~Werner, T.~Ohgoe, N.~Prokof'ev and B.~Svistunov,
\newblock \emph{{Diagrammatic Monte Carlo algorithm for the resonant Fermi
  gas}},
\newblock Phys. Rev. B \textbf{99}, 035140 (2019),
\newblock \doi{10.1103/PhysRevB.99.035140}.

\bibitem{ProkofevSvistunovBoldPRL}
N.~Prokof'ev and B.~Svistunov,
\newblock \emph{Bold Diagrammatic Monte Carlo Technique: When the Sign Problem
  Is Welcome},
\newblock Phys. Rev. Lett. \textbf{99}, 250201 (2007),
\newblock \doi{10.1103/PhysRevLett.99.250201}.

\bibitem{RossiCDet}
R.~Rossi,
\newblock \emph{Determinant Diagrammatic Monte Carlo in the Thermodynamic
  Limit},
\newblock Phys. Rev. Lett. \textbf{119}, 045701 (2017),
\newblock \doi{10.1103/PhysRevLett.119.045701}.

\bibitem{SimkovicCDet}
F.~\ifmmode~\check{S}\else \v{S}\fi{}imkovic and E.~Kozik,
\newblock \emph{Determinant Monte Carlo for irreducible Feynman diagrams in the
  strongly correlated regime},
\newblock Phys. Rev. B \textbf{100}, 121102(R) (2019),
\newblock \doi{10.1103/PhysRevB.100.121102}.

\bibitem{MoutenetCDet}
A.~Moutenet, W.~Wu and M.~Ferrero,
\newblock \emph{Determinant Monte Carlo algorithms for dynamical quantities in
  fermionic systems},
\newblock Phys. Rev. B \textbf{97}, 085117 (2018),
\newblock \doi{10.1103/PhysRevB.97.085117}.

\bibitem{SimkovicCrossover}
F.~\ifmmode~\check{S}\else \v{S}\fi{}imkovic, J.~P.~F. LeBlanc, A.~J. Kim,
  Y.~Deng, N.~V. Prokof'ev, B.~V. Svistunov and E.~Kozik,
\newblock \emph{Extended Crossover from a Fermi Liquid to a
  Quasiantiferromagnet in the Half-Filled 2D Hubbard Model},
\newblock Phys. Rev. Lett. \textbf{124}, 017003 (2020),
\newblock \doi{10.1103/PhysRevLett.124.017003}.

\bibitem{KimKozik2020}
A.~J. Kim, F.~Simkovic and E.~Kozik,
\newblock \emph{Spin and Charge Correlations across the Metal-to-Insulator
  Crossover in the Half-Filled 2D Hubbard Model},
\newblock Phys. Rev. Lett. \textbf{124}, 117602 (2020),
\newblock \doi{10.1103/PhysRevLett.124.117602}.

\bibitem{JohanInfiniteU}
J.~Carlstr\"om,
\newblock \emph{Diagrammatic Monte Carlo procedure for the spin-charge
  transformed Hubbard model},
\newblock Phys. Rev. B \textbf{97}, 075119 (2018),
\newblock \doi{10.1103/PhysRevB.97.075119}.

\bibitem{KunHauleEG}
K.~Chen and K.~Haule,
\newblock \emph{A combined variational and diagrammatic quantum Monte Carlo
  approach to the many-electron problem},
\newblock Nature Comm. \textbf{10}, 3725 (2019),
\newblock \doi{10.1038/s41467-019-11708-6}.

\bibitem{HauleChenEG_2}
K.~Haule and K.~Chen,
\newblock \emph{Single-particle excitations in the uniform electron gas by
  diagrammatic Monte Carlo},
\newblock Sci. Rep. \textbf{12}(1), 2294 (2022),
\newblock \doi{10.1038/s41598-022-06188-6}.

\bibitem{GullPRR2020}
J.~Li, M.~Wallerberger and E.~Gull,
\newblock \emph{Diagrammatic Monte Carlo method for impurity models with
  general interactions and hybridizations},
\newblock Phys. Rev. Research \textbf{2}, 033211 (2020),
\newblock \doi{10.1103/PhysRevResearch.2.033211}.

\bibitem{SimkovicLadderCDet}
F.~\ifmmode~\check{S}\else \v{S}\fi{}imkovic, R.~Rossi and M.~Ferrero,
\newblock \emph{Efficient one-loop-renormalized vertex expansions with
  connected determinant diagrammatic Monte Carlo},
\newblock Phys. Rev. B \textbf{102}, 195122 (2020),
\newblock \doi{10.1103/PhysRevB.102.195122}.

\bibitem{RossiRDet}
R.~Rossi, F.~{\v{S}}imkovic and M.~Ferrero,
\newblock \emph{Renormalized perturbation theory at large expansion orders},
\newblock Europhys. Lett. \textbf{132}, 11001 (2020),
\newblock \doi{10.1209/0295-5075/132/11001}.

\bibitem{SimkovicSuscept_k}
F.~{\ifmmode \check{S}\else \v{S}\fi{}imkovic}, R.~Rossi and M.~Ferrero,
\newblock \emph{Two-dimensional Hubbard model at finite temperature: Weak,
  strong, and long correlation regimes},
\newblock Phys. Rev. Res. \textbf{4}, 043201 (2022),
\newblock \doi{10.1103/PhysRevResearch.4.043201}.

\bibitem{LenihanTc}
C.~Lenihan, A.~J. Kim, F.~\ifmmode~\check{S}\else \v{S}\fi{}imkovic and
  E.~Kozik,
\newblock \emph{Evaluating Second-Order Phase Transitions with Diagrammatic
  Monte Carlo: N\'eel Transition in the Doped Three-Dimensional Hubbard Model},
\newblock Phys. Rev. Lett. \textbf{129}, 107202 (2022),
\newblock \doi{10.1103/PhysRevLett.129.107202}.

\bibitem{SpadaSF}
G.~Spada, R.~Rossi, F.~Simkovic, R.~Garioud, M.~Ferrero, K.~{Van Houcke} and
  F.~Werner,
\newblock \emph{{High-order expansion around BCS theory}},
\newblock \href{https://arxiv.org/abs/2103.12038}{arXiv:2103.12038}.

\bibitem{SimkovicPG}
F.~Simkovic, R.~Rossi, A.~Georges and M.~Ferrero,
\newblock \href{https://arxiv.org/abs/2209.09237}{arXiv:2209.09237}.

\bibitem{WietekTriang}
A.~Wietek, R.~Rossi, F.~\ifmmode~\check{S}\else \v{S}\fi{}imkovic, M.~Klett,
  P.~Hansmann, M.~Ferrero, E.~M. Stoudenmire, T.~Sch\"afer and A.~Georges,
\newblock \emph{Mott Insulating States with Competing Orders in the Triangular
  Lattice Hubbard Model},
\newblock Phys. Rev. X \textbf{11}, 041013 (2021),
\newblock \doi{10.1103/PhysRevX.11.041013}.

\bibitem{ChenWDM}
P.-C. Hou, B.-Z. Wang, K.~Haule, Y.~Deng and K.~Chen,
\newblock \emph{Exchange-correlation effect in the charge response of a warm
  dense electron gas},
\newblock Phys. Rev. B \textbf{106}, L081126 (2022),
\newblock \doi{10.1103/PhysRevB.106.L081126}.

\bibitem{TupitsynRealFreq}
I.~S. Tupitsyn, A.~M. Tsvelik, R.~M. Konik and N.~V. Prokof'ev,
\newblock \emph{Real-Frequency Response Functions at Finite Temperature},
\newblock Phys. Rev. Lett. \textbf{127}, 026403 (2021),
\newblock \doi{10.1103/PhysRevLett.127.026403}.

\bibitem{LeblancTupitsynEG}
J.~P.~F. LeBlanc, K.~Chen, K.~Haule, N.~V. Prokof'ev and I.~S. Tupitsyn,
\newblock \emph{Dynamic Response of an Electron Gas: Towards the Exact
  Exchange-Correlation Kernel},
\newblock Phys. Rev. Lett. \textbf{129}, 246401 (2022),
\newblock \doi{10.1103/PhysRevLett.129.246401}.

\bibitem{TupitsynLandauDamping}
I.~S. Tupitsyn and N.~V. Prokof'ev,
\newblock \emph{Landau Damping in an Electron Gas},
\newblock \href{https://arxiv.org/abs/2311.05611}{arXiv:2311.05611}.

\bibitem{KozikCombinatorialSummation}
E.~Kozik,
\newblock \emph{Combinatorial summation of Feynman diagrams: Equation of state
  of the 2D SU(N) Hubbard model},
\newblock \href{https://arxiv.org/abs/2309.13774}{arXiv:2309.13774}.

\bibitem{RossiNesting}
R.~Rossi, F.~Simkovic, M.~Ferrero, A.~Georges, A.~M. Tsvelik, N.~V. Prokof'ev
  and I.~S. Tupitsyn,
\newblock \emph{Interaction-enhanced nesting in Spin-Fermion and Fermi-Hubbard
  models},
\newblock \href{https://arxiv.org/abs/2402.13238}{arXiv:2402.13238}.

\bibitem{CarlstromStrongCouplingDiagMC}
J.~Carlstr\"om,
\newblock \emph{Strong-coupling diagrammatic Monte Carlo technique for
  correlated fermions and frustrated spins},
\newblock Phys. Rev. B \textbf{103}, 195147 (2021),
\newblock \doi{10.1103/PhysRevB.103.195147}.

\bibitem{CarlstromTMD}
J.~Carlstr\"om,
\newblock \emph{In situ controllable magnetic phases in doped twisted bilayer
  transition metal dichalcogenides},
\newblock Phys. Rev. Res. \textbf{4}, 043126 (2022),
\newblock \doi{10.1103/PhysRevResearch.4.043126}.

\bibitem{CarlstromSpectralTopol}
J.~Carlstr\"om,
\newblock \emph{Spectral topology and its relation to Fermi arcs in strongly
  correlated systems},
\newblock Phys. Rev. Res. \textbf{5}, 033160 (2023),
\newblock \doi{10.1103/PhysRevResearch.5.033160}.

\bibitem{KimStrangeMetal}
A.~J. Kim, P.~Werner and E.~Kozik,
\newblock \emph{Strange Metal Solution in the Diagrammatic Theory for the $2d$
  Hubbard Model},
\newblock \href{https://arxiv.org/abs/2012.06159}{arXiv:2012.06159}.

\bibitem{InPrepFFS}
R. Rossi {\it et al.}, to be published.

\end{thebibliography}
\let\addcontentsline\oldaddcontentsline

\end{document}